# Radiation of a charge moving at an angle to the face of a dielectric prism

**Sergey N. Galyamin and Andrey V. Tyukhtin**

## Annotation

Cherenkov radiation which generates by a charge moving in presence of a dielectric prism from the base to the top is analyzed. Unlike our previous works, here we consider the case when the charge trajectory is not parallel to the prism face. We apply a new version of the aperture method which uses the expansion in terms of plane waves in the prism material and on the aperture. The integrals for the radiation field are calculated analytically in the Fraunhofer region where the field components are written as single integrals. The influence of the incline of the charge trajectory on the radiation characteristics is discussed.

## Introduction

Radiation of charged particles moving in presence of dielectric objects (``targets") of complicated form is of interests for various applications [1-5]. As an example, one can mention a new method of bunch diagnostics which requires calculation of Cherenkov radiation outside dielectric object [2]. Typically, the size of the target is much larger than the wavelengths under consideration. On the one hand, this fact complicates considerably computer simulations because very large amount of resources and time is required. However, on the other hand, this fact gives us an obvious small parameter of the problem and allows developing approximate methods of analysis.

We developed two methods for solution of such problems which can be named ``ray-optic method" and ``aperture method". Both these methods are valid for objects which are much larger than the wavelengths under consideration. The basis of the methods and solutions of different problems with use of them are described in series of our works (the last of them are the papers [6-12]). We will not repeat here the main steps of these methods. Note only that the aperture method is more general than the ray optics one (the last one cannot describe different diffraction effects and requires additional limitation on the observation point place). Therefore we will use here the aperture technique.

It should be emphasized that this method has been tested for various objects, in particular, for a dielectric cone and a ball with vacuum channel (see [6, 10] and the references given there). Testing was done by comparison with COMSOL simulations. As a result, it has been shown (as expected) that the aperture method gives an accuracy of the order of the ratio of the considered wavelength to the size of the object (at least in the regions of the maximal field magnitudes).

In this paper, we consider promising target in the shape of a prism. This object is actively used not only in optics but also in accelerator and beam physics. For instance, such a target is offered to be applied in the new noninvasive method of bunch diagnostics [2,5].

Here we consider the case of charge moving from the base of the prism to its top. Similar problem was considered by us in the paper [] where we assumed that the charge moves strictly parallel to the face of the prism. However this assumption is hard for implementation in the experiment. Therefore, it is important to know how the inclination of the charge trajectory with respect to the prism face affects the excited radiation.

In this paper, it is assumed that the charge trajectory makes some nonzero (but sufficiently small) angle with respect to the prism face. This geometry greatly complicates the solution of the problem. Therefore, we use here a new version of the aperture method. We use the fact that the target has only plane borders. This allows using the expansion in the plane waves only. Unlike our previous works we will not analytically calculate the integrals which determine the field components inside the prism. The integrals will be calculated only for the radiation field outside the prism in the Fraunhofer region.

# 1. Aperture integrals: general form and approximation for Fraunhofer zone

We remind the Straton-Chu formulas (known also as "aperture integrals") in a form convenient for our researches [6-10]. According to them the Fourier transform of electric field can be written in the following general form (we use Gaussian system of units):

$$\vec{E}\left(\vec{R}\right) = \vec{E}^{(h)}\left(\vec{R}\right) + \vec{E}^{(e)}\left(\vec{R}\right),$$
$$\vec{E}^{(h)}\left(\vec{R}\right) = \frac{ik}{4\pi}\int_{\Sigma}\left\{\left[\vec{n}'\times\vec{H}\left(\vec{R}'\right)\right]G\left(\left|\vec{R}-\vec{R}'\right|\right) + \frac{1}{k^2}\left(\left[\vec{n}'\times\vec{H}\left(\vec{R}'\right)\right]\cdot\nabla'\right)\nabla' G\left(\left|\vec{R}-\vec{R}'\right|\right)\right\}d\Sigma', \quad (1)$$
$$\vec{E}^{(e)}\left(\vec{R}\right) = \frac{1}{4\pi}\int_{\Sigma}\left[\left[\vec{n}'\times\vec{E}\left(\vec{R}'\right)\right]\times\nabla' G\left(\left|\vec{R}-\vec{R}'\right|\right)\right]d\Sigma',$$

where $\Sigma$ is an aperture area, $\vec{E}\left(\vec{R}'\right)$, $\vec{H}\left(\vec{R}'\right)$ is the field on the aperture, $k = \omega/c$ is a wave number of the outer space (which is vacuum), $\vec{n}'$ is a unit external normal to the aperture in the point $\vec{R}'$ (it is directed into the area where the observation point is placed), $G(R) = \exp(i\,kR)/R$ is a Green function of Helmholtz equation, and $\nabla'$ is a gradient: $\nabla' = \vec{e}_x\,\partial/\partial x' + \vec{e}_y\,\partial/\partial y' + \vec{e}_z\,\partial/\partial z'$. Analogous formulas are known for the magnetic field as well, however we do not write them here because we are interested in a "wave" zone $kR >> 1$ ($R$ is a distance from the aperture to the observation point) where $\left|\vec{E}\right| \approx \left|\vec{H}\right|$.

The observation point is often located far from the target, in other words, in the region where so-called "wave parameter" $D$ is large:

$$D \sim \lambda R/\Sigma \sim \lambda R/d^2 >> 1, \quad (2)$$

where $\lambda = 2\pi/k$ is a wavelength under consideration, and $\Sigma \sim d^2$ is a square of an aperture. This region (which usually called the Fraunhofer area, or far-field area) is the main interest for the present work. The condition (2) automatically results in the inequality

$$R >> d\cdot(d/\lambda) >> d\,, \quad (3)$$

because $d >> \lambda$ in the problem under consideration. Using the inequalities (2), (3) one can obtain the following approximate formulas for Fraunhofer area:

$$\vec{E}^{(h)}\left(\vec{R}\right) \approx \frac{ik\exp(ikR)}{4\pi R}\int_{\Sigma}\left\{\left[\vec{n}'\times\vec{H}\left(\vec{R}'\right)\right] - \vec{e}_R\left(\vec{e}_R\cdot\left[\vec{n}'\times\vec{H}\left(\vec{R}'\right)\right]\right)\right\}\exp\left(-ik\vec{e}_R\vec{R}'\right)d\Sigma',$$
$$\vec{E}^{(e)}\left(\vec{R}\right) \approx \frac{ik\exp(ikR)}{4\pi R}\int_{\Sigma}\left[\vec{e}_R\times\left[\vec{n}'\times\vec{E}\left(\vec{R}'\right)\right]\right]\exp\left(-ik\vec{e}_R\vec{R}'\right)d\Sigma', \quad (4)$$

where $\vec{e}_R = \vec{R}/R$.

## 2. Solution of the “key” problem in the form of expansion on plane waves

The geometry of the problem under consideration is shown in Fig.1. The point charge $q$ moves from the prism base to its top at the angle $\alpha$ with respect to the “lower” face of the prism (the $z$-axis is oriented along this face). The prism has the apex angle $\alpha_p$. The prism material (being isotropic, homogeneous, and nonconductive) are characterized by permittivity $\varepsilon$, permeability $\mu$, and the refractive index $n=\sqrt{\varepsilon\mu}$. It is assumed that these values do not depend on the wave vector (the spatial dispersion is absent) but they can be depend on the frequency $\omega$ (the frequency dispersion can be taken into account). The medium surrounding the prism is a vacuum.

The charge velocity is $\vec{V}=c\vec{\beta}=c\beta\sin\alpha\cdot\vec{e}_y+c\beta\cos\alpha\cdot\vec{e}_z$. The charge and current densities are determined correspondingly by the expressions $\rho=q\delta(x)\delta(y-c\beta\sin\alpha\cdot t)\delta(z-c\beta\cos\alpha\cdot t)$, $\vec{j}=\vec{V}\rho$, where $\delta(\xi)$ is the Dirac delta-function.

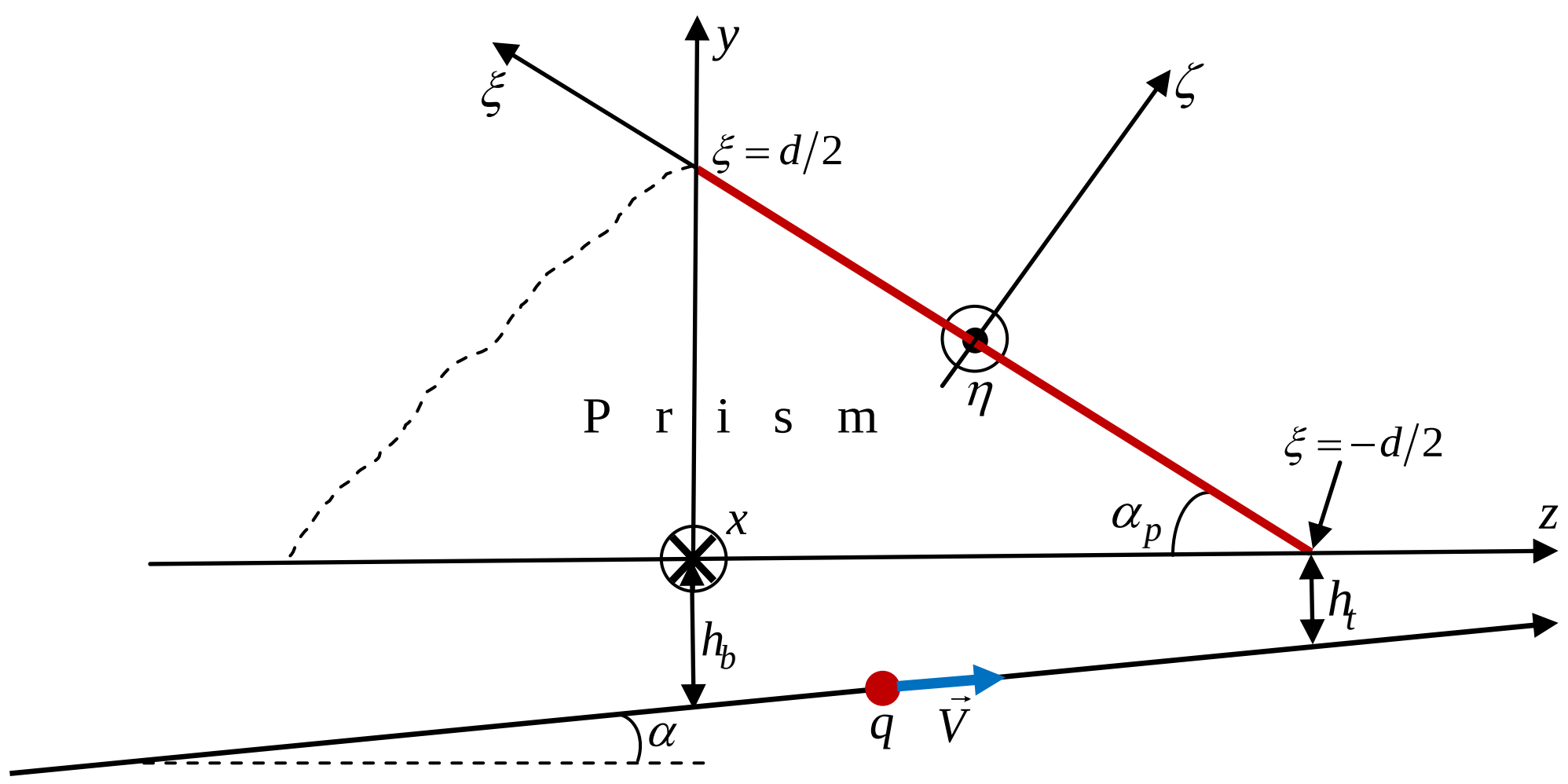

Fig.1. The longitudinal section of the prism and the charge trajectory.

We assume that the charge velocity is more than the phase velocity at the considered frequency $\omega$, so that Cherenkov radiation is generated in the prism material. This radiation falls on the external (oblique) face of the prism and exits the prism through the rectangular area which plays the role of the “aperture” (the bold red line in Fig.1). Note that, along with coordinates $x, y, z$, we will use coordinates $\xi, \eta, \zeta$: $\zeta=0$ is the plane of the aperture and the point $\xi=\eta=\zeta=0$ is the aperture centre. In these coordinates, the aperture is determined by the inequalities $|\xi|<d/2$, $|\eta|<b/2$. Note that the shape of the "rear" face of the target (dotted line in Fig. 1) does not matter. The only important thing is that the aperture is completely illuminated by Cherenkov radiation.

Note that the values $\vec{E}$ and $\vec{H}$ given further are the Fourier transforms of the electric and magnetic fields. The fields themselves are determined by Fourier integrals in the form $\int\limits_{-\infty}^{\infty}\vec{E}e^{-i\omega t}d\omega$.

We will write out formulas for positive frequencies only ( $\omega > 0$ ). Expressions for negative frequencies are easily obtained by the rule $\vec{E}(-\omega) = \vec{E}^{*}(\omega)$ (asterisk means complex conjugation) that follows from the reality of the field components.

The method applied by us uses the solution of certain "key" problem. In the case under consideration, it is the problem about the field of the charge moving along the boundary $y = 0$ of the dielectric half-space $y > 0$ and the vacuum half-space $y < 0$. It should be noted that, in the problem under consideration, there is a significant difference from those considered by us earlier: in the "key problem", the charge trajectory crosses the boundary of the half-space occupied by the dielectric. However, we are interested in the case of a relatively small trajectory slope: $|\alpha| << 1$. Therefore, we can assume that the charge crosses the plane $y = 0$ at the distance of many wavelengths from the front (at $\alpha > 0$ ) or rear (at $\alpha < 0$ ) vertices of the prism. Under such conditions, Cherenkov radiation is excited practically only when the charge passes near the lower (see Fig. 1) face.

Earlier we have developed and applied such a variant of the aperture method, in which it was supposed to find asymptotics of the field inside the target (which is possible due to its large size). However, in a geometrically complex problem, this path is not optimal. Here we will not calculate the asymptotics of the field inside the prism, but will restrict ourselves to finding the expansion in plane waves. As will be shown below, this eventually leads to integrals for the field components outside the prism in a form that is quite convenient in the most interesting Fraunhofer region.

We do not give here the cumbersome analytical calculations of the field inside the prism. We give only the results which are necessary for following: the expressions for the field components on the outer surface of the aperture in the form of an expansion in plane waves. These expressions have the following form:

$$\begin{Bmatrix} E_\xi^a \\ E_h^a \\ H_\xi^a \\ H_\eta^a \end{Bmatrix} = \frac{q\omega}{2\pi c^2} \int_{-\infty}^{\infty} dk_x \begin{Bmatrix} W_e^{(||)} k_\zeta^{(t1)} k_\xi^{(i1)} - W_e^{(\perp)} k_0 k_x \\ W_e^{(||)} k_x k_\zeta^{(t1)} + W_e^{(\perp)} k_0 k_\xi^{(i1)} \\ -W_h^{(||)} k_0 k_x + W_h^{(\perp)} k_\zeta^{(t1)} k_\xi^{(i1)} \\ W_h^{(||)} k_0 k_\xi^{(i1)} + W_h^{(\perp)} k_x k_\zeta^{(t1)} \end{Bmatrix} \times$$

$$\times \frac{1}{a_1 k_0} \exp\left\{ i k_y^{(t)} \frac{d}{2} \tan\alpha_p + i k_{z\alpha} \frac{d}{2} + i k_{y\alpha} h_b \right\} \exp\left\{ i k_\xi^{(i1)} \xi + i k_x \eta \right\}, \tag{5}$$

where $k_0 = \omega/c$,

$$W_e^{(||)} = T_{||1} \left\{ f(k_x)\left(\vec{A}\cdot\vec{A}_1\right) + \frac{g(k_x)}{\mu}\left(\left[\frac{\vec{k}^{(t)}}{k_0}\times\vec{A}\right]\cdot\vec{A}_1\right) \right\},$$

$$W_e^{(\perp)} = T_{\perp 1} \left\{ -\frac{f(k_x)}{\varepsilon}\left(\left[\frac{\vec{k}^{(t)}}{k_0}\times\vec{A}\right]\cdot\vec{A}_1\right) + g(k_x)\left(\vec{A}\cdot\vec{A}_1\right) \right\},$$

$$W_h^{(||)} = T_{||1} \left\{ f(k_x)\left(\vec{A}\cdot\vec{A}_1\right) + \frac{g(k_x)}{\mu}\left(\left[\frac{\vec{k}^{(t)}}{k_0}\times\vec{A}\right]\cdot\vec{A}_1\right) \right\}, \tag{6}$$

$$W_h^{(||)} = T_{\perp 1} \left\{ \frac{f(k_x)}{\varepsilon}\left(\left[\frac{\vec{k}^{(t)}}{k_0}\times\vec{A}\right]\cdot\vec{A}_1\right) - g(k_x)\left(\vec{A}\cdot\vec{A}_1\right) \right\};$$

$$f(k_x)=T_{\|}\frac{k_{z\alpha}k_{y\alpha}\left[k_0\left(\beta^{-2}-1\right)\cos^2\alpha-\beta^{-1}k_{y0}\sin\alpha\right]+k_0k_x^2\sin\alpha}{a\,k_{y0}\left(k_{z\alpha}^2-k_0^2\right)},$$
$$g(k_x)=T_{\perp}k_x\frac{k_{z\alpha}k_{y\alpha}\sin\alpha-k_0^2\left(\beta^{-2}-1\right)\cos^2\alpha+\beta^{-1}k_0k_{y0}\sin\alpha}{a\,k_{y0}\left(k_{z\alpha}^2-k_0^2\right)}. \quad (7)$$

We use here the following designations:

$$T_{\|}=\frac{2\varepsilon k_{y\alpha}}{\varepsilon k_{y\alpha}+k_y^{(t)}},\qquad T_{\perp}=\frac{2\mu k_{y\alpha}}{\mu k_{y\alpha}+k_y^{(t)}}. \quad (8)$$

$$T_{\|1}=\frac{2k_\zeta^{(i1)}}{k_\zeta^{(i1)}+\varepsilon k_\zeta^{(t1)}},\qquad T_{\perp1}=\frac{2k_\zeta^{(i1)}}{k_\zeta^{(i1)}+\mu k_\zeta^{(t1)}}. \quad (9)$$

$$\vec{A}=\frac{-k_{z\alpha}\vec{e}_x+k_x\vec{e}_z}{a},\qquad a=\sqrt{k_x^2+k_{z\alpha}^2}\,. \quad (10)$$

$$\vec{A}_1=\frac{1}{a_1}\left\{-k_x\vec{e}_\xi+k_\xi^{(i1)}\vec{e}_\eta\right\},\qquad a_1=\sqrt{k_x^2+\left(k_\xi^{(i1)}\right)^2}. \quad (11)$$

$$\begin{aligned}
&\vec{k}^{(t)}=\vec{k}^{(i1)}=k_x\vec{e}_x+k_y^{(t)}\vec{e}_y+k_{z\alpha}\vec{e}_z,\qquad k_y^{(t)}=\sqrt{k_0^2\varepsilon\mu-k_x^2-k_{z\alpha}^2},\\
&k_{z\alpha}=\beta^{-1}k_0\cos\alpha-k_{y0}\sin\alpha,\qquad k_{y\alpha}=k_{y0}\cos\alpha+\beta^{-1}k_0\sin\alpha,\\
&k_{y0}=i\sqrt{k_x^2+k_0^2\left(\beta^{-2}-1\right)},\\
&\vec{k}^{(t1)}=a_1\vec{\tau}+k_\zeta^{(t1)}\vec{e}_\zeta,\qquad \vec{\tau}=\frac{1}{a_1}\left\{k_x\vec{e}_x+k_\xi^{(i1)}\sin\alpha_p\vec{e}_y-k_\xi^{(i1)}\cos\alpha_p\vec{e}_z\right\},\\
&k_\xi^{(i1)}=k_y^{(t)}\sin\alpha_p-k_{z\alpha}\cos\alpha_p,\qquad k_\zeta^{(i1)}=k_y^{(t)}\cos\alpha_p+k_{z\alpha}\sin\alpha_p,\\
&k_\zeta^{(t1)}=\sqrt{k_0^2\left(1-\varepsilon\mu\right)+\left(k_\zeta^{(i1)}\right)^2}=\sqrt{k_0^2-k_x^2-\left(k_\xi^{(i1)}\right)^2}.
\end{aligned} \quad (12)$$

The physical meaning of some notations is as follows: $T_{\|}$ and $T_{\perp}$ are transmission coefficients of waves of vertical ($\|$) and horizontal ($\perp$) polarizations at the boundary $y=0$; $T_{\|1}$ and $T_{\perp1}$ are corresponding transmission coefficients at the aperture; $\vec{A}$ is a unity vector orthogonal to the plane of incidence on the boundary $y=0$; $\vec{A}_1$ is a unity vector orthogonal to the plane of incidence on the aperture. Note that the parameters $h_b$ and $h_t$ are related by the relation

$$h_b=h_t+d\cos\alpha_p\,\mathrm{tg}\alpha\,. \quad (13)$$

## 3. Analytical results for Fraunhofer (far-field) area

The aperture integrals for prismatic object are written in a convenient form in the paper [7]. In the Fraunhofer area, it is convenient to use the spherical coordinates $R,\Theta,\Phi$ which are associated with coordinates $\xi,\eta,\zeta$ :

$$\xi=R\sin\Theta\cos\Phi,\quad \eta=R\sin\Theta\sin\Phi,\quad \zeta=R\cos\Theta. \quad (14)$$

Using these coordinates one can obtain from (7) the following expressions [7]:

$$E_R = 0,$$

$$\begin{Bmatrix} E_\Theta \\ E_\Phi \end{Bmatrix} \approx \frac{ik\exp(ikR)}{4\pi R}\int_\Sigma \begin{Bmatrix} -E_\xi^a\cos\Phi - E_\eta^a\sin\Phi + \left[H_\xi^a\sin\Phi - H_\eta^a\cos\Phi\right]\cos\Theta \\ H_\xi^a\cos\Phi + H_\eta^a\sin\Phi + \left[E_\xi^a\sin\Phi - E_\eta^a\cos\Phi\right]\cos\Theta \end{Bmatrix}\times$$
$$\times\exp\left\{-ik\left(\xi'\cos\Phi + \eta'\sin\Phi\right)\sin\Theta\right\}d\Sigma', \tag{15}$$

where $E_{\xi,\eta}^a = E_{\xi,\eta}^a(\vec{R}')$, $H_{\xi,\eta}^a = H_{\xi,\eta}^a(\vec{R}')$, and $\Sigma$ is an aperture area. The first of the equations (15) means that the wave is transversal (as expected). Naturally, the magnetic field value is equal to electric field value, and vectors $\vec{E},\vec{H},\vec{R}$ form the right orthogonal triad.

Substituting (5) into (15), after some transformations and changing the order of integration, we obtain the following result:

$$\begin{Bmatrix} E_\Theta \\ E_\Phi \end{Bmatrix} \approx \frac{iq\omega^2\exp(ikR)}{8\pi^2 c^3 R}\int_{-\infty}^{\infty}\frac{F}{k_0 a_1}\begin{Bmatrix} U_\Theta \\ U_\Phi \end{Bmatrix}\exp\left\{ik_y^{(t)}\frac{d}{2}\tan\alpha_p + ik_{z\alpha}\frac{d}{2} + ik_{y\alpha}h\right\}dk_x\,, \tag{16}$$

where

$$\begin{aligned} U_\Theta &= -\left(W_e^{(\|)}k_\zeta^{(t1)} + W_h^{(\|)}k_0\cos\Theta\right)\left(k_x\sin\Phi + k_\xi^{(i1)}\cos\Phi\right) + \\ &\quad + \left(W_e^{(\perp)}k_0 - W_h^{(\perp)}k_\zeta^{(t1)}\cos\Theta\right)\left(k_x\cos\Phi - k_\xi^{(i1)}\sin\Phi\right), \\ U_\Phi &= \left(W_h^{(\|)}k_0 + W_e^{(\|)}k_\zeta^{(t1)}\cos\Theta\right)\left(k_\xi^{(i1)}\sin\Phi - k_x\cos\Phi\right) + \\ &\quad + \left(W_h^{(\perp)}k_\zeta^{(t1)} - W_e^{(\perp)}k_0\cos\Theta\right)\left(k_\xi^{(i1)}\cos\Phi + k_x\sin\Phi\right); \end{aligned} \tag{17}$$

$$\begin{aligned} F &= \int_{-d/2}^{d/2}d\xi'\int_{-b/2}^{b/2}d\eta'\exp\left\{i\left(k_\xi^{(i1)} - k\cos\Phi\sin\Theta\right)\xi' + i\left(k_x - k\sin\Phi\sin\Theta\right)\eta'\right\} = \\ &= 4\frac{\sin\left[\left(k_\xi^{(i1)} - k\cos\Phi\sin\Theta\right)d/2\right]}{k_\xi^{(i1)} - k\cos\Phi\sin\Theta}\frac{\sin\left[\left(k_x - k\sin\Phi\sin\Theta\right)b/2\right]}{k_x - k\sin\Phi\sin\Theta}. \end{aligned} \tag{18}$$

As one can see, due to the fact that the inner integral $F$ in (16) is easily found, calculation of the Fourier transforms of the field components is reduced to calculating the single integrals. Of course, this result refers only to the Fraunhofer region, since, for arbitrary distances, the integrals over $\xi'$, $\eta'$ are much more complicated [7].

# 4. Numerical results and discussion

The energy passing through the unit square is $W = \int_{-\infty}^{\infty} S dt$, where $\vec{S}$ is Pointing vector: $\vec{S} = \frac{c}{4\pi}\left[\int_{-\infty}^{\infty}\vec{E}e^{-i\omega t}d\omega\times\int_{-\infty}^{\infty}\vec{H}e^{-i\omega t}d\omega\right]$. One can show that the value $W$ can be written in the form $W = \int_0^\infty w\,d\omega$, where $w = c\left|\vec{E}\right|^2 = c\left(\left|E_\Theta\right|^2 + \left|E_\Phi\right|^2\right)$ is a spectral density of an energy which passes through the unity square for the entire time. For convenience, we give further the plots for the value $R\left|\vec{E}\right| = R\sqrt{w/c}$ which is do not depend on the distance $R$.

Further we use the angle $\tilde{\Theta}$ which is equal to

$$\tilde{\Theta} = \begin{cases} \Theta \ \text{ for } \ 0 \le \Phi < \pi, \\ -\Theta \ \text{ for } \ \pi \le \Phi < 2\pi. \end{cases} \quad (19)$$

The use of this angle is convenient for demonstration of the angle dependence for sections $\Phi = \Phi_0$ and $\Phi = \Phi_0 + \pi$ on one graphics. Figures 2, 3 show the dependences of $R|\vec{E}|$ on the angle $\tilde{\Theta}$. Figure 2 refers to the case $\beta = 0.8$, and Fig. 3 refers to the case of ultra relativistic charge motion ($\beta \approx 1$). Other parameters are the same for both figures. The distance from the charge trajectory to the top of the prism is taken the same for all curves ($h_t = 2k_0^{-1}$). Different curves in each figure correspond to different angles of inclination of the trajectory: the red solid curves refer to motion parallel to the face ($\alpha = 0$), the blue dashed curves refer to the case when the charge approaches the prism ($\alpha < 0$), and the green dotted curves refer to the case when the charge moves away from the prism ($\alpha > 0$). Note that since the distance $h_t$ is fix, then the trajectory is "on average" closer to the prism in the second case than in the first and the third.

In the case of $\beta = 0.8$ (Fig. 2), the main maximum lies in the plane of symmetry ($y = 0$): the maximum field value in the upper plot is much larger than in the lower one. As one can see, its position changes very little with slight change in $\alpha$: the difference in the maximum's positions in the upper plot (Fig. 2) does not exceed $0.1^0$. In the same time, the field magnitude changes very significantly. This is explained by the fact that the approach of the charge trajectory to the prism leads to significant increase in the excitation of radiation.

The last regularity is true for the main maximums in the case of ultrarelativistic charge motion also (Fig. 3). However, the essential difference is that, in this case, the main maximums do not lie in the plane of symmetry, but are significantly shifted from it (the maximum values of the field in the lower plot are much larger than in the upper one). This circumstance is explained by the fact that, for $V \approx c$, the radiation maximums in the prism material lie at certain distance from the symmetry plane [13]. Note as well that the shift of the main maximums with change of $\alpha$ in the ultrarelativistic case is essential: when $\alpha$ changes from $2^0$ to $-2^0$, the shifts of maximums are approximately $8^0$.

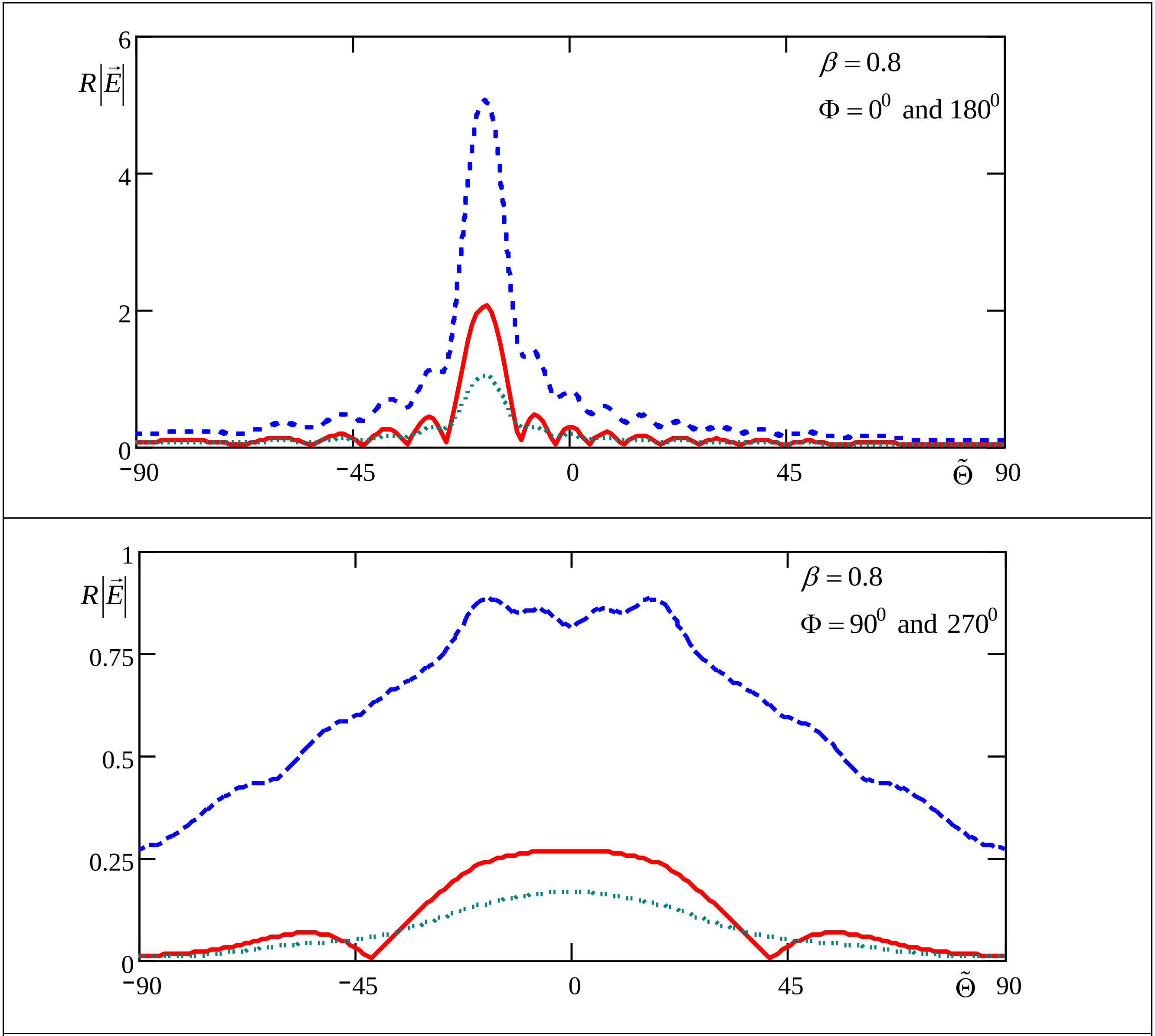

Fig. 2. Dependence of the value $R|\vec{E}|$ (in unities $q/c$) on the angle $\tilde{\Theta}$ in two sections: $\Phi = 0, 180^0$ (top) and $\Phi = 90^0, 270^0$ (bottom). Parameters: $\varepsilon = 4$, $\mu = 1$, $d = b = 50 \cdot k_0^{-1}$, $\alpha_p = 30^0$, $\beta = 0.8$, $h_t = 2 \cdot k_0^{-1}$; $\alpha = 0$ (red solid line), $\alpha = -2^0$ (dashed blue line), $\alpha = 2^0$ (green dotted line).

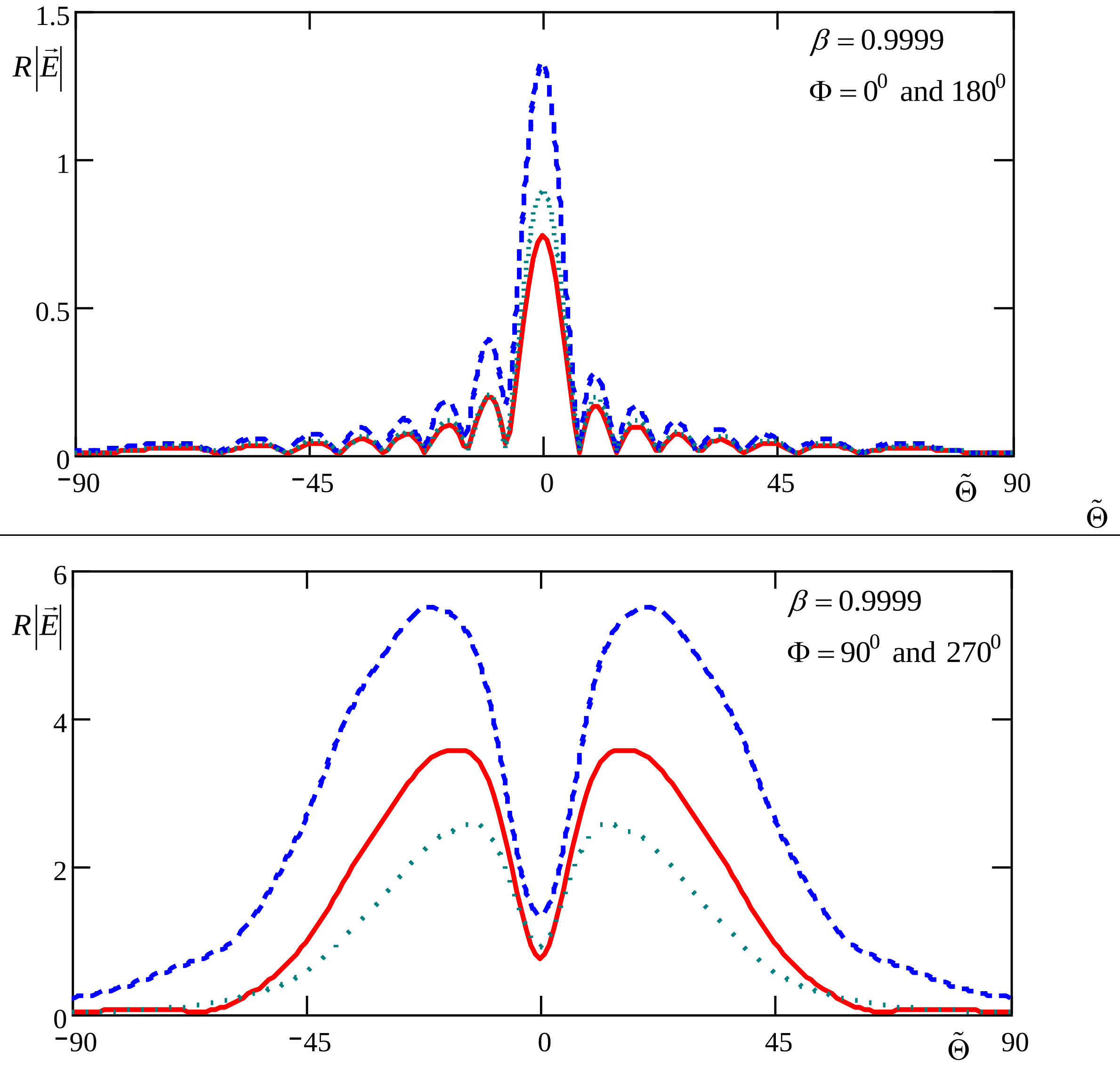

Fig. 3. Dependence of the value $R|\vec{E}|$ (in unities $q/c$) on the angle $\tilde{\Theta}$ in two sections: $\Phi = 0, 180^0$ (top) and $\Phi = 90^0, 270^0$ (bottom). Parameters: $\varepsilon = 4$, $\mu = 1$, $d = b = 50 \cdot k_0^{-1}$, $\alpha_p = 30^0$, $\beta = 0.9999$, $h_t = 2 \cdot k_0^{-1}$; $\alpha = 0$ (red solid line), $\alpha = -2^0$ (dashed blue line), $\alpha = 2^0$ (green dotted line).

## Conclusion

Cherenkov radiation generated by a charge moving in the presence of a dielectric prism was analyzed. Unlike our previous works, we considered the case when the charge trajectory is not parallel to the prism face. This geometry greatly complicates the solution of the problem. Therefore we applied the new version of the aperture method, which uses the expansion in terms of plane waves in the prism material and on the aperture. The integrals for the radiation field were calculated in the Fraunhofer region. As a result, the Fourier transforms of the field components have been written as single integrals which were used further for the numerical calculation.

It has been shown, in particular, that the field magnitude depends significantly on the inclination of the charge trajectory with respect to the prism face. The positions of the main maximums are much more stable. The dependence of the positions of the main maximums on the trajectory inclination angle is more significant for the ultrarelativistic charge motion compared to the cases of a lower charge velocity.

## Acknowledgment

This work was supported by the Russian Science Foundation (Grant No. 18-72-10137).